\documentclass[11pt]{article}
\usepackage{fullpage, times, amssymb}

\newtheorem{fact}{Fact}[section]    
\newtheorem{theorem}{Theorem}[section]    
\newcommand{\qed}{\hfill{$\rule{6pt}{6pt}$}} 
\newenvironment{proof}{\noindent{\bf Proof}:}{\qed}

\newcommand{\R}{{\mathbb R}}

\newcommand{\E}{{\mathsf{E}}}

\newcommand{\Tr}{{\mathsf{Tr}}}

\newcommand{\trnorm}[1]{\left\| #1 \right\|_{{\mathrm{tr}}}}

\title{Distinguishing sets of quantum states}
\author{
Rahul Jain \\
U.C. Berkeley \\
{\sf rahulj@cs.berkeley.edu}
\thanks{
This work was supported by an Army Research Office (ARO), North
California,  grant number DAAD 19-03-1-00082.
}
}
\date{}

\begin{document}

\maketitle

\begin{abstract}
Given two sets finite $S_0$ and $S_1$ of quantum states. We show necessary and
sufficient conditions for distinguishing them by a measurement.
\end{abstract}
Let there be two finite sets of quantum states $S_0 = \{\rho_i : 1 \leq i
\leq l \}$ and $S_1 = \{\sigma_i : 1 \leq i \leq l \}$. Please
see~\cite{nielsen:quant} for a good introduction to information
theory. By a {\em $\epsilon$-separating measurement} $T$ we mean a
POVM element $T$ such that $\forall \rho \in S_0$ and $\forall \sigma
\in S_1$ we have $\Tr T \rho - \Tr T \sigma \geq \epsilon$ for some
constant $\epsilon$. For distributions $\mu_0 $ on $S_0$ and $\mu_1$
on $S_1$, let $\rho_{\mu_0} = \E_{i \in \mu_0} [\rho_i]$ and $\sigma_{i
\in \mu_1} = \E_{\mu_1}[ \sigma_i]$.  We show the following:

\begin{theorem}
The following statements are equivalent:
\begin{enumerate}
\item 
There exists a $\epsilon$-separating measurement $T$.
\item For all distributions $\mu_0$ on $S_0$ and $\mu_1$ on $S_1$,
$\trnorm{\rho_{\mu_0} - \sigma_{\mu_1}} \geq 2\epsilon$.
\end{enumerate}
\end{theorem}
For our proof we will need the following facts.
\begin{fact}[~\cite{nielsen:quant}]
\label{fact:dist}
Given quantum states $\rho, \sigma$,
$$ \max_{T: \mathrm{~a~POVM~element}} \Tr T \rho - \Tr T \sigma  =
\frac{1}{2} \trnorm{\rho - \sigma} $$
\end{fact}
We have the following minimax theorem from game
theory(see~\cite{osborne:gametheory}).
\begin{fact}
\label{fact:minimax}
Let $A_1,A_2$ be non-empty, either finite or convex and compact
subsets of $\R^n$. Let $u: A_1 \times A_2 \rightarrow \R$ be a
continuous function. Let $\mu_1, \mu_2$ be distributions on $A_1$ and
$A_2$ respectively. Then,
\begin{displaymath}
\min_{\mu_1}\, \max_{a_2\in A_2} E_{\mu_1}[u(a_1,a_2)] 
= \max_{\mu_2}\, \min_{a_1 \in A_1} E_{\mu_2}[u(a_1,a_2)]
\end{displaymath}
\end{fact}
\begin{proof} \\
\noindent
1) $\Rightarrow$ 2) : If $T$ is a separating measurement then from
linearity of $\Tr$ operation we see that for all distributions $\mu_0$ on $S_0$
and $\mu_1$ on $S_1$, $\Tr T
\rho_{\mu_0} - \Tr T \sigma_{\mu_1} \geq \epsilon$. Fact~\ref{fact:dist} now implies $\trnorm{\rho_{\mu_0} - \sigma_{\mu_1}} \geq 2\epsilon$.
\\ \\
\noindent
2) $\Rightarrow$ 1): Let us define sets $A_1 =
\{T : T \textrm{ a POVM element} \}$ and $A_2 =
\{(\rho, \sigma) : \rho \in S_0, \sigma \in S_1 \}$. Let the function $u:
A_1 \times A_2 \rightarrow \R$ be defined as $u(T, (\rho,\sigma)) =
\Tr T \rho - \Tr T \sigma$. For a distribution $\mu$ on $S_0 \times
S_1$, let $\mu_0$ be the marginal distribution on $S_0$ and $\mu_1$ be
the marginal distribution on $S_1$. From fact~\ref{fact:minimax} and
the fact that a convex combination of POVM elements is also a POVM
element and from linearity of $\Tr$ operation it follows:
\begin{eqnarray*}
 \max_T \min_{(\rho, \sigma) \in S_0 \times S_1} \Tr T \rho - \Tr T
\sigma  & = &  \min_{\mu} \max_T \E_{\mu}[\Tr T \rho - \Tr T \sigma]
\\
& = & \min_{\mu} \max_T \Tr T \rho_{\mu_0} - \Tr T
\sigma_{\mu_1} \\
& = & \min_{\mu} \frac{1}{2} \trnorm{\rho_{\mu_0} - \sigma_{\mu_1}} (\textrm{from Fact~\ref{fact:dist}})
\end{eqnarray*}
From above it is clear that 2) $\Rightarrow$ 1).
\end{proof}
\\
\noindent
{\bf \large Acknowledgment:} We thank Julia Kempe, Rohit Khandekar
and Oded Regev for useful discussions and patiently listening to an
early proof.

\end{document}